\documentclass[prl,twocolumn,preprintnumbers,superscriptaddress,amsmath,amssymb]{revtex4-1}
 \usepackage[dvips,final]{graphicx}
  \usepackage{amssymb}
   \usepackage{amsmath}
    \usepackage{amsfonts}
     \usepackage{epsfig}
      \usepackage{bm}

\usepackage[section]{placeins}

\usepackage{sidecap}
\usepackage{multirow}
\usepackage{booktabs}
\usepackage{array}
\usepackage{tabularx}
\usepackage{xcolor}
\usepackage{pstricks}

\begin{document}

\newcommand{\ds}{\displaystyle}
\newcommand{\mc}{\multicolumn} 
\newcommand{\bce}{\begin{center}}
\newcommand{\ece}{\end{center}}
\newcommand{\beq}{\begin{equation}}
\newcommand{\eeq}{\end{equation}}
\newcommand{\bea}{\begin{eqnarray}}

\newcommand{\eea}{\end{eqnarray}}
\newcommand{\cont}{\nonumber\eea\bea}
\newcommand{\cl}[1]{\begin{center} {#1} \end{center}}
\newcommand{\ea}{\end{array}}

\newcommand{\ab}{{\alpha\beta}}
\newcommand{\cd}{{\gamma\delta}}
\newcommand{\dc}{{\delta\gamma}}
\newcommand{\ac}{{\alpha\gamma}}
\newcommand{\bd}{{\beta\delta}}
\newcommand{\abc}{{\alpha\beta\gamma}}
\newcommand{\eps}{{\epsilon}}
\newcommand{\lam}{{\lambda}}
\newcommand{\mn}{{\mu\nu}}
\newcommand{\mpnp}{{\mu'\nu'}}
\newcommand{\Amuu}{{A_{\mu}}}
\newcommand{\Amuo}{{A^{\mu}}}
\newcommand{\Vmuu}{{V_{\mu}}}
\newcommand{\Vmuo}{{V^{\mu}}}
\newcommand{\Anuu}{{A_{\nu}}}
\newcommand{\Anuo}{{A^{\nu}}}
\newcommand{\Vnuu}{{V_{\nu}}}
\newcommand{\Vnuo}{{V^{\nu}}}
\newcommand{\Fmnu}{{F_{\mu\nu}}}
\newcommand{\Fmno}{{F^{\mu\nu}}}

\newcommand{\abcd}{{\alpha\beta\gamma\delta}}


\newcommand{\bsigma}{\mbox{\boldmath $\sigma$}}
\newcommand{\beps}{\mbox{\boldmath $\varepsilon$}}
\newcommand{\btau}{\mbox{\boldmath $\tau$}}
\newcommand{\brho}{\mbox{\boldmath $\rho$}}
\newcommand{\bpipi}{\mbox{\boldmath $\pi\pi$}} 
\newcommand{\bss}{\bsigma\!\cdot\!\bsigma}
\newcommand{\btt}{\btau\!\cdot\!\btau}
\newcommand{\bnabla}{\mbox{\boldmath $\nabla$}}
\newcommand{\bphi}{\mbox{\boldmath $\tau$}}
\newcommand{\bvarphi}{\mbox{\boldmath $\rho$}}
\newcommand{\bE}{\mbox{\boldmath $E$}}
\newcommand{\bDelta}{\mbox{\boldmath $\Delta$}}
\newcommand{\bGamma}{\mbox{\boldmath $\Gamma$}}
\newcommand{\bpsi}{\mbox{\boldmath $\psi$}}
\newcommand{\bPsi}{\mbox{\boldmath $\Psi$}}
\newcommand{\bPhi}{\mbox{\boldmath $\Phi$}}
\newcommand{\bnab}{\mbox{\boldmath $\nabla$}}
\newcommand{\bpi}{\mbox{\boldmath $\pi$}}
\newcommand{\btheta}{\mbox{\boldmath $\theta$}}
\newcommand{\bkappa}{\mbox{\boldmath $\kappa$}}
\newcommand{\bgamma}{\mbox{\boldmath $\gamma$}}

\newcommand{\bp}{\mbox{\boldmath $p$}}
\newcommand{\ba}{\mbox{\boldmath $a$}}
\newcommand{\bq}{\mbox{\boldmath $q$}}
\newcommand{\br}{\mbox{\boldmath $r$}}
\newcommand{\bs}{\mbox{\boldmath $s$}}
\newcommand{\bk}{\mbox{\boldmath $k$}}
\newcommand{\bl}{\mbox{\boldmath $l$}}
\newcommand{\bb}{\mbox{\boldmath $b$}}
\newcommand{\be}{\mbox{\boldmath $e$}}
\newcommand{\bP}{\mbox{\boldmath $P$}}
\newcommand{\bV}{\mbox{\boldmath $V$}}
\newcommand{\bI}{\mbox{\boldmath $I$}}
\newcommand{\bJ}{\mbox{\boldmath $J$}}

\newcommand{\bT}{{\bf T}}
\newcommand{\fph}{${\cal F}$}
\newcommand{\aph}{${\cal A}$}
\newcommand{\dph}{${\cal D}$}
\newcommand{\fpi}{f_\pi}
\newcommand{\mpi}{m_\pi}
\newcommand{\Tr}{{\mbox{\rm Tr}}}
\def\Qb{\overline{Q}}
\newcommand{\delu}{\partial_{\mu}}
\newcommand{\delo}{\partial^{\mu}}
\newcommand{\up}{\!\uparrow}
\newcommand{\upup}{\uparrow\uparrow}
\newcommand{\updo}{\uparrow\downarrow}
\newcommand{\uu}{$\uparrow\uparrow$}
\newcommand{\ud}{$\uparrow\downarrow$}
\newcommand{\auu}{$a^{\uparrow\uparrow}$}
\newcommand{\aud}{$a^{\uparrow\downarrow}$}
\newcommand{\pu}{p\!\uparrow}
\newcommand{\qp}{quasiparticle}
\newcommand{\sa}{scattering amplitude}
\newcommand{\ph}{particle-hole}
\newcommand{\qcd}{{\it QCD}}
\newcommand{\integ}{\int\!d}
\newcommand{\ie}{{\sl i.e.~}}
\newcommand{\etal}{{\sl et al.~}}
\newcommand{\etc}{{\sl etc.~}}
\newcommand{\rhs}{{\sl rhs~}}
\newcommand{\lhs}{{\sl lhs~}}
\newcommand{\eg}{{\sl e.g.~}}
\newcommand{\ef}{\epsilon_F}
\newcommand{\sigt}{d^2\sigma/d\Omega dE}
\newcommand{\sige}{{d^2\sigma\over d\Omega dE}}
\newcommand{\rpaeq}{\beq
\left ( \begin{array}{cc}
A&B\\
-B^*&-A^*\end{array}\right )
\left ( \begin{array}{c}
X^{(\kappa})\\Y^{(\kappa)}\end{array}\right )=E_\kappa
\left ( \begin{array}{c}
X^{(\kappa})\\Y^{(\kappa)}\end{array}\right )
\eeq}

\newcommand{\ket}[1]{{#1} \rangle}
\newcommand{\bra}[1]{\langle {#1} }

\newcommand{\Bigket}[1]{{#1} \Big\rangle}
\newcommand{\Bigbra}[1]{\Big\langle {#1} }

\newcommand{\ave}[1]{\langle {#1} \rangle}
\newcommand{\Bigave}[1]{\left\langle {#1} \right\rangle}
\newcommand{\half}{{1\over 2}}

\newcommand{\singlespace}{
    \renewcommand{\baselinestretch}{1}\large\normalsize}
\newcommand{\doublespace}{
    \renewcommand{\baselinestretch}{1.6}\large\normalsize}
\newcommand{\bftau}{\mbox{\boldmath $\tau$}}
\newcommand{\bfalpha}{\mbox{\boldmath $\alpha$}}
\newcommand{\bfgamma}{\mbox{\boldmath $\gamma$}}
\newcommand{\bfxi}{\mbox{\boldmath $\xi$}}
\newcommand{\bfbeta}{\mbox{\boldmath $\beta$}}
\newcommand{\bfeta}{\mbox{\boldmath $\eta$}}
\newcommand{\bfpi}{\mbox{\boldmath $\pi$}}
\newcommand{\bfphi}{\mbox{\boldmath $\phi$}}
\newcommand{\bfR}{\mbox{\boldmath ${\cal R}$}}
\newcommand{\bfL}{\mbox{\boldmath ${\cal L}$}}
\newcommand{\bfM}{\mbox{\boldmath ${\cal M}$}}
\def\dblint{\mathop{\rlap{\hbox{$\displaystyle\!\int\!\!\!\!\!\int$}}
    \hbox{$\bigcirc$}}}
\def\ut#1{$\underline{\smash{\vphantom{y}\hbox{#1}}}$}

\def\UNITY{{\bf 1\! |}}
\def\Pom{{\bf I\!P}}
\def\lsim{\mathrel{\rlap{\lower4pt\hbox{\hskip1pt$\sim$}}
    \raise1pt\hbox{$<$}}}         
\def\gsim{\mathrel{\rlap{\lower4pt\hbox{\hskip1pt$\sim$}}
    \raise1pt\hbox{$>$}}}         

\newcommand\scalemath[2]{\scalebox{#1}{\mbox{\ensuremath{\displaystyle #2}}}}

\newcommand{\RP}[1]{{\blue RP: #1}}
\newcommand{\WS}[1]{{\red WS: #1}}

\title{Probing the structure of $\chi_{c1}(3872)$
with photon transition form factors}

\author{Izabela Babiarz}
\email{izabela.babiarz@ifj.edu.pl}
\affiliation{Institute of Nuclear Physics, Polish Academy of Sciences, 
ul. Radzikowskiego 152, PL-31-342 Krak{\'o}w, Poland}

\author{Roman Pasechnik}
\email{roman.pasechnik@hep.lu.se}
\affiliation{Department of Physics,
Lund University, SE-223 62 Lund, Sweden}

\author{Wolfgang Sch\"afer}%
\email{wolfgang.schafer@ifj.edu.pl}
\affiliation{Institute of Nuclear
Physics, Polish Academy of Sciences, ul. Radzikowskiego 152, PL-31-342 
Krak{\'o}w, Poland}

\author{Antoni Szczurek}
\email{antoni.szczurek@ifj.edu.pl}
\affiliation{Institute of Nuclear
Physics, Polish Academy of Sciences, ul. Radzikowskiego 152, PL-31-342 
Krak{\'o}w, Poland}
\affiliation{College of Mathematics and Natural Sciences,
University of Rzesz\'ow, ul. Pigonia 1, PL-35-310 Rzesz\'ow, Poland\vspace{5mm}}

\begin{abstract}
We propose to study the structure of the enigmatic $\chi_{c1}(3872)$ axial vector meson through its $\gamma^*_L \gamma \to \chi_{c1}(3872)$ transition form factor. We derive a light-front wave function representation of the form factor for the lowest $c \bar c$ Fock-state. We found that the reduced width of the state is well within the current experimental bound recently published by the Belle collaboration. This strongly suggests a crucial role of the $c \bar c$ Fock-state in the photon-induced production. Our results for the $Q^2$ dependence can be tested by future single tagged $e^+ e^-$ experiments, giving further insights into the short-distance structure of this meson. 
\end{abstract}

\pacs{12.38.Bx, 13.85.Ni, 14.40.Pq}
\maketitle

\textit{Introduction}--Starting from the discovery of the $\chi_{c1}(3872)$ state by the Belle collaboration \cite{Belle:2003nnu}, recent years have seen a surge of discoveries of new hadronic states \cite{Lebed:2016hpi, Olsen:2017bmm, Chen:2022asf}. The microscopic structure of $\chi_{c1}(3872)$ state is still under intense debate in the literature \cite{Karliner:2017qhf,Brambilla:2019esw}. While its quantum numbers $J^{\rm PC} = 1^{++}$ imply a possible $\chi_{c1}(2P)$ charmonium \cite{Achasov:2015oia}, its mass appears to be close to the $D^0 \bar D^{*0}$ threshold suggesting a picture based on a very weakly bound meson molecule \cite{Tornqvist:2004qy} (for a review, see e.g.~Ref.~\cite{Guo:2017jvc} and references therein). The latter option also gives a natural explanation of the strong isospin violation in its decays. However, such a violation can also be accommodated in the structure of couplings to $c \bar c$ and meson-meson channels~\cite{Danilkin:2010cc,Ferretti:2013faa,Bruschini:2022bsh,Coito:2012vf,Yamaguchi:2019vea}, e.g.~via mixing of molecular and compact states studied in Ref.~\cite{Kang:2016jxw}. For a discussion of another possible picture of $\chi_{c1}(3872)$ as a compact tetraquark, see Refs.~\cite{Maiani:2004vq, Ali:2017jda}.

There is an open question of to what extent production mechanisms of $\chi_{c1}(3872)$ can shed light on its internal structure. The inclusive $\chi_{c1}(3872)$ production cross section has been recently measured at the LHC as a function of its transverse momentum \cite{ATLAS:2016kwu,CMS:2013fpt,LHCb:2021ten}. A striking similarity with the production rate of $\psi(2S)$ points to the importance of a compact component in the wave function. Indeed, the measured transverse momentum distributions are well described assuming a large $c \bar c$ component \cite{Cisek:2022uqx}. Studying $\chi_{c1}(3872)$ production in a cleaner environment than hadronic collisions would provide new opportunities to understand its structure better. 

In this Letter, we work under the basic assumption of a $c \bar c$ $2P$-state and suggest probing the $c \bar c$ component through the production of $\chi_{c1}(3872)$ in the (virtual) photon-photon channel in single-tagged $e^+ e^-$ collisions. A first attempt to perform such a measurement has been made by the Belle collaboration providing a bound on the reduced width~\cite{Belle:2020ndp}. A recent result of Ref.~\cite{Li:2021ejv} found the reduced width significantly overshooting the Belle bound, hence, concluding that the $c \bar c$ component is not relevant for the structure of $\chi_{c1}(3872)$ as probed by photons. This represents a controversy with recent studies in hadronic reactions \cite{Cisek:2022uqx}, where the $c\bar c$ component almost explains experimental data.

\begin{figure}
    \centering
    \includegraphics[width=0.4\textwidth]{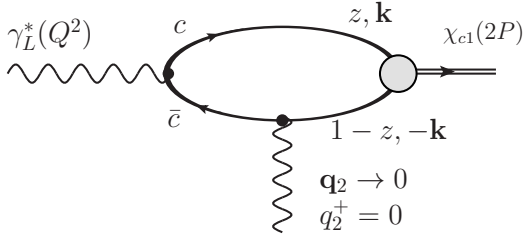}
    \caption{An illustration of the meson production mechanism in photon-photon fusion in the light-cone dipole picture, with relevant ingredients and kinematics. }
    \label{fig:diagram}
\end{figure}

\textit{Transition form factor}--Due to the Landau-Yang theorem, at least one off-shell photon is required for $\chi_{c1}$ production in the photon-photon fusion channel. Here, we study the corresponding form factor for one longitudinal and one transverse photon representing the amplitude of a longitudinal photon to an axial $c\bar c$ meson transition in an external electromagnetic field approaching the forward amplitude. For this purpose, we utilize the Light-Front (LF) approach to transition form factors in the Drell-Yan frame \cite{Brodsky:1997de}. Here, the longitudinal photon with spacelike virtuality $Q_1^2 \equiv - q_1^2$ carries four-momentum $q_1 = (q_{1+}, q_{1-}, \bf{0}_\perp)$, with $q_{1-} = - Q_1^2 /(2 q_{1+})$ and polarization vector $\varepsilon_L = 1/Q_1\, (q_{1+},-q_{1-}, \bf{0}_\perp)$, while the plus-momentum of the second photon vanishes $q_2^+ = 0$, such that $q_2 = (0, q_{2-}, \bq_2)$. In the real photon limit, $Q_2^2 \equiv -q_2^2 = \bq_2^2 \to 0$, i.e.~when its transverse momentum $\bq_2\to 0$, the transition amplitude vanishes linearly with $\bq_2$ enabling us to extract the relevant form factor. Indeed, in the considered frame, the LF plus component of the electromagnetic current is free from parton number changing and instantaneous fermion exchange contributions \cite{Brodsky:1998hn}, and can be written in terms of the lowest $c \bar c$ Fock-state LF wave functions (LFWFs) (from now on, we define $Q^2 \equiv Q_1^2$):
\begin{multline}    
    \bra{\chi_{c1}(\lambda_A)}| J_+(0) |\ket{\gamma^*_L(Q^2)}
    = 2 q_{1+} \, \sqrt{N_c}\int \frac{dz d^2\bk} {z(1-z) 16 \pi^3} \\ 
    \times \sum_{\lambda, \bar \lambda} \Psi^{(\lambda_A)*}_{\lambda \bar \lambda} (z, \bk)
    \,  (\bq_2 \cdot \nabla_{\bk}) \Psi^{\gamma_L}_{\lambda \bar \lambda}(z, \bk, Q^2)\, .
    \label{eq:J_plus_WF}
\end{multline}
Above, the summation over the (anti)quark color indices has been performed, $N_c=3$ is the number of colors in QCD, and we introduced the LF helicity $\lambda_A = \pm 1,0$ of the axial meson $\chi_{c1}$, as well as $c\bar c \to \chi_{c1}$ and $\gamma^*_L \to c\bar c$ LFWFs, $\Psi^{(\lambda_A)*}_{\lambda \bar \lambda}$ and $\Psi^{\gamma_L}_{\lambda \bar \lambda}$, respectively. The integration is over the internal LF momenta of quark $(c)$ and antiquark $(\bar c)$, namely, the LF momentum fraction $z = k_{c+}/q_{1+}$ of the $c$-quark and its transverse momentum $\bk$ as illustrated in Fig.~\ref{fig:diagram}. The (anti)quark coupling to the external field conserves the LF helicities $\lambda, \bar \lambda$ of quark and antiquark, whose $\pm 1/2$ values are denoted by $\uparrow$ and $\downarrow$, respectively. 
Furthermore, it is instructive to utilize the general covariant parametrization of the $\gamma^* \gamma^* \to \chi_{c1}$ amplitude of Ref.~\cite{Babiarz:2022xxm}, which is similar to the one found in Ref.~\cite{Poppe:1986dq} and is based on $\gamma^* \gamma^*$ c.m.~frame helicity amplitudes. We notice that only one term contributes to the transition amplitude of interest in the limit $Q_2^2 \to 0$,
\begin{multline}
\varepsilon_L^\mu n^{-\nu} {\mathcal M}_{\mu \nu \rho} E^{*\rho} \rightarrow 
4 \pi \alpha_{\rm em}\, \tilde G_{\nu \rho} n_-^\nu E^{*\rho} \, 
\frac{F_{\rm LT}(Q^2,0)}{q_1\cdot q_2} \, ,
\end{multline}
where $\alpha_{\rm em} = e^2/4\pi$ is the fine structure constant, $E=E(\lambda_A)$ is the polarization vector of the axial meson, $n_-= (0,1,\bf{0}_\perp)$, and $\tilde G_{\nu \rho} \equiv \varepsilon_{\nu \rho \alpha \beta} q_1^\alpha q_2^\beta$. We choose the LF spin projection $\lambda_A = +1$, and obtain:
\begin{multline}    
     \bra{\chi_{c1}(+1)}| J_+(0) |\ket{\gamma^*_L(Q^2)} =  \,  2 q_{1+} \, \frac{q_{2x} - i q_{2y}}{\sqrt{2}} \\ 
     \times \frac {\sqrt{4 \pi \alpha_{\rm em}} F_{\rm LT}(Q^2,0)}{Q^2 + M_{\chi}^2} \, ,
     \label{eq:J_plus_FF}
\end{multline}
in terms of the considered $\chi_{c1}$ meson mass $M_{\chi}=(3871.65\pm 0.06)$~MeV \cite{ParticleDataGroup:2022pth} and the photon-meson transition form factor $F_{\rm LT}(Q^2,0)$. Combining this expression with Eq.~(\ref{eq:J_plus_WF}) and using the well-known expression for the perturbative LF wave function of the longitudinal photon's $c \bar c$ component (see e.g.~Ref.~\cite{Kovchegov:2012mbw}),
\begin{eqnarray}
   \Psi^{\gamma_L}_{\lambda \bar \lambda}(z, \bk, Q^2) = 
   e e_c \sqrt{z(1-z)}\, {2 z (1-z) Q \over \bk^2 + \epsilon^2} \, \delta_{\lambda, -\bar \lambda}\,, 
\end{eqnarray}
with $\epsilon^2 = m_c^2 + z(1-z) Q^2$, charm (anti)quark mass $m_c$, and the electric charge of the charm quark is $e_c=2/3$, one arrives at the LFWF representation of the transition form factor:
\begin{multline}    
\frac{f_{\rm LT}(Q^2)}{Q^2 + M_\chi^2} = -2\sqrt{2 N_c} \,  e_c
\int \frac{dz d^2\bk}{16 \pi^3} \, \frac{k_x + i k_y}{[\bk^2 + \epsilon^2]^2}   \\
\times  \sqrt{z(1-z)} \Big\{ \Psi^{(+1)*}_{\uparrow \downarrow}(z,\bk) + 
\Psi^{(+1)*}_{\downarrow \uparrow}(z,\bk) \Big\} \, .
\label{eq:fLT}
\end{multline}
The dimensionless transition from factor
$f_{\rm LT}(Q^2) \equiv F_{\rm LT}(Q^2,0)/Q$ takes a finite value in the limit $Q^2 \to 0$.
The representation of Eq.~(\ref{eq:fLT}) can also be derived from more general expressions for the transition form factors for two spacelike virtual photons found  earlier in Ref.~\cite{Babiarz:2022xxm}.
\begin{figure}[h!]
    \centering
    \includegraphics[width = 0.45\textwidth]{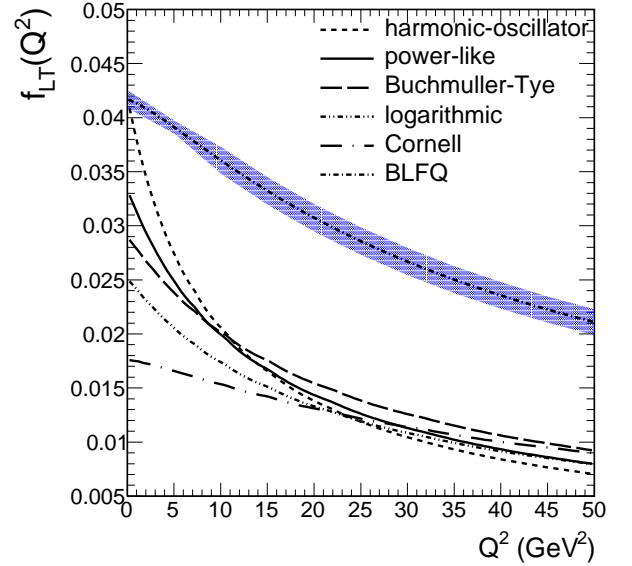}
    \caption{The dimensionless $\gamma^*_L \gamma \to \chi_{c1}(2P)$ transition form factor $f_{\rm LT}(Q^2)$ found in Eq.~(\ref{eq:fLT}). }
    \label{fig:TransitionFF}
\end{figure}

A brief comment on the considered LFWFs is in order. In our analysis, we adopt two different approaches. The first one is based on $c \bar c$ radial wave functions in the $c \bar c$-pair rest frame obtained by solving the Schr\"odinger equation for a variety of phenomenologically viable potential models. Then, one appropriately transforms both the resulting radial wave functions \cite{Terentev:1976jk}, and their spin-orbital components \cite{Melosh:1974cu} in order to describe the boosted meson states in the Drell-Yan frame. For further details of this procedure, an overview of the potential models and related theoretical uncertainties, see Refs.~\cite{Krelina:2018hmt, Cepila:2019skb} and \cite{Babiarz:2022xxm} for photoproduction of vector and axial mesons, respectively.  In the considered case of the first radial excitation of the $\chi_{c1}$ meson -- the $2P$ state -- the relevant combination of LFWFs takes the form:
\begin{multline}
\sqrt{z(1-z)} \, \Big\{ \Psi^{(+1)*}_{\uparrow \downarrow}(z,\bk) +   \Psi^{(+1)*}_{\downarrow \uparrow}(z,\bk) \Big\} \\ = (k_x - i k_y) \, \sqrt{\frac{3}{2}} \frac{\pi \sqrt{M_{c \bar c}}}{2} \frac{u_{\rm 2P}(k)}{k^2} \, ,
\end{multline}
where $u_{\rm 2P}(k)$ is the radial wave function in the $c \bar c$-pair rest frame found for a given interquark interaction potential as a function of the relative $c$ and $\bar c$ three-momentum, $k = \sqrt{M_{c \bar c}^2 - 4 m_c^2}/2$ \cite{Babiarz:2022xxm}, with $M^2_{c \bar c} = (\bk^2 + m_c^2)/z(1-z)$. The second approach is based on the Basis Light Front Quantization (BLFQ) of Refs.~\cite{Li:2015zda,Li:2017mlw,Li:2021ejv}, where the LFWFs are obtained without referring to non-relativistic $c \bar c$ interaction potentials and factorisation of radial and spin-orbit components applicable only in the $c \bar c$-pair rest frame. 
Instead, one formulates an LF-Hamiltonian problem
\begin{eqnarray}
    H_{\rm eff} |\ket{\chi_c; \lambda_A, P_+, \bP} = M^2 |\ket{\chi_c; \lambda_A, P_+, \bP} \, , 
\end{eqnarray}
where the effective Hamiltonian used in \cite{Li:2015zda,Li:2017mlw} contains a term motivated by a ``soft-wall'' confinement from LF-holography, as well as a longitudinal confinement potential supplemented by one gluon exchange including the full spin-structure. The LFWFs from the Fock state decomposition
\begin{multline}
   |\ket{\chi_c; \lambda_A ,P_+, \bP} = \sum_{\lambda \bar \lambda} \int \frac{dz d^2 \bk}{z(1-z)16 \pi^3} \, \psi_{\lambda \bar \lambda}^{(\lambda_A)}(z,\bk) \\
   \times
   \frac{\delta^i_j}{\sqrt{N_c}} |\ket {c_i(z,\bk + z \bP) \bar c^j(1-z,-\bk+(1-z) \bP)}
\end{multline}
are then obtained by discretizing the Hamiltonian and evaluating it on a finite basis. 
In our calculations, we have used the LFWFs from numerical results of \cite{Li:2017mlw} published in Ref.~\cite{Li:2017MenData}.

\textit{Numerical results}--In Fig.~\ref{fig:TransitionFF} we show our results for the dimensionless transition from factor $f_{\rm LT}(Q^2) \equiv F_{\rm LT}(Q^2,0)/Q$ over a broad range of $Q^2$. We see considerable dependence on the potential model at $Q^2 \lsim 15 \,\rm{GeV}^2$. Here the BLFQ approach stands out from the nonrelativistic potential approach. 
It exhibits a much weaker dependence on $Q^2$. We trace that partially to the broader distribution in LF momentum fraction $z$ exhibited by the BLFQ-WF.
The blue band in Fig.~\ref{fig:TransitionFF} was obtained in a similar manner as in \cite{Li:2015zda}. For selected $Q^2$ values, we calculated $f_{LT}(Q^2)$ for different values of the number of basis functions $N_{\rm basis}$.
Then $f_{LT}(Q^2)$ was obtained by extrapolating $f_{LT}(Q^2,N_{\rm basis})$ for $1/N_{\rm basis} \to 0$ by fitting polynomials of different order to points for different $N_{\rm basis}$.
The error band reflects the spread due to the order of the extrapolating polynomial. The precision of reduced width ${\tilde \Gamma}_{\gamma \gamma}$ is better than 2 \%.

Now, let us remind the reader of the definition of the so-called reduced $\gamma\gamma$ decay width of $\chi_{c1}$ given in the limit of the vanishing projectile photon virtuality \cite{TPCTwoGamma:1988izb}:
\begin{eqnarray}    
\tilde \Gamma_{\gamma\gamma} &=& \lim_{Q^2 \to 0} \frac{M_{\chi}^2}{Q^2} \Gamma^{\rm LT}_{\gamma^* \gamma^*} (Q^2,0,M_{\chi}^2)
\nonumber \\
&=& \frac{ \pi \alpha^2_{\rm em} M_{\chi}}{3} \, f^2_{\rm LT}(0) \,.
\end{eqnarray}
Hence, a measurement of the reduced width provides direct access to the dimensionless transition form factor in the limit $Q^2 \to 0$. We summarize our results for $f_{\rm LT}(0)$ and the reduced width in Table~\ref{tab:red_widths}. We observe a considerable spread between the results for different potential models. Here the Cornell potential gives the smallest value of the reduced width, roughly a factor of five smaller than the value obtained for the harmonic oscillator potential. Notice that the quark mass substantially influences the normalization of the form factor $f_{\rm LT}(0)$. 
The BLFQ-WF gives the largest result among the considered approaches. However, it is almost a factor of six smaller than the result obtained by Li et al.~\cite{Li:2021ejv} using the same wave functions. These authors, though, do not use the plus-component of the current and derive a different representation of the transition form factor.
\begin{table}
    \caption{The reduced width of the $\chi_{c1}(2P)$ state for several models of the charmonium wave functions with specific $c$-quark mass.}
    \centering
    \begin{ruledtabular}
    \begin{tabular}{lccc}
    $c\bar c$ potential & $m_c$ (GeV) & $f_{\rm LT}(0)$ & $\tilde \Gamma_{\gamma\gamma}$ (keV) \\
    \colrule
    harmonic oscillator& 1.4   & 0.041 & 0.36\\
    power-law          & 1.334 & 0.033 & 0.24\\
    Buchm\"uller-Tye   & 1.48  & 0.029 & 0.18\\
    logarithmic        & 1.5   & 0.025 & 0.14\\
    Cornell            & 1.84  & 0.018 & 0.07\\
    BLFQ \cite{Li:2017MenData} & 1.6 & 0.044 & 0.42\\
    \end{tabular}
    \end{ruledtabular}
    \label{tab:red_widths}
\end{table}

The Belle collaboration has reported the first evidence for the production of $\chi_{c1}(3872)$ in single-tag $e^+ e^-$ collisions \cite{Belle:2020ndp}.
From three measured events, they provided a range for its reduced width,
$0.02\,\rm{keV} < \tilde{\Gamma}_{\gamma\gamma} < 0.5\,\rm{keV}$. This result has recently been updated by Achasov et al. \cite{Achasov:2022puf} using a corrected value for the branching ratio ${\rm Br}(\chi_{c1}(3872)\to \pi^+ \pi^- J/\psi)$ \cite{ParticleDataGroup:2022pth} and reads
\begin{eqnarray}
 0.024\,\rm{keV} < \tilde{\Gamma}_{\gamma\gamma}(\chi_{c1}(3872)) < 0.615\,\rm{keV}
\end{eqnarray}
Using nonrelativistic quark model relations, Achasov et al. \cite{Achasov:2022puf} provided the following estimate
\begin{eqnarray}
 \tilde{\Gamma}_{\gamma\gamma}(\chi_{c1}(3872)) \approx 0.35\,\rm{keV}\, \div 0.93 \, \rm{keV} \, .
\end{eqnarray}
Even with the large dependence on the $c\bar c$ potential, all our results, including the BLFQ approach, lie well within the experimentally allowed range. Therefore, $\gamma \gamma$ data do not exclude the $c \bar c$ 
option, although there is certainly some room for a contribution from an additional meson-meson component.
Regarding the molecular scenario, no estimates for the reduced width in the molecular scenario are available.

\textit{Summary and outlook}--In this work, we concentrated on the lowest $c \bar c$ Fock state consistent with the quantum numbers of the axial vector $\chi_{c1}(3872)$. There are many indications that meson-meson components in the wave function may also be necessary to understand the 
decay properties of $\chi_{c1}(3872)$. 
The transition amplitude of Eq.~(\ref{eq:J_plus_WF})
can be understood as a transition dipole moment
\begin{multline}    
    \bra{\chi_{c1}(\lambda_A)}| J_+(0) |\ket{\gamma^*_L(Q^2)} \\
    \propto i \bq_2 \cdot  \bra{\chi_{c1}(\lambda_A)}| \br  |\ket{\gamma^*_L(Q^2)} \, .
\end{multline}
Evidently, the electric dipoles which enter this transition are controlled by the photon LFWF to have sizes  ${r \sim 1/\epsilon \sim (m_c^2 + Q^2/4)}^{-1/2} \lesssim 0.15 \, \rm{fm}$.
This is much smaller than the expected size of a molecular component, whose wave function extends up to distances of around $10 \, \rm{fm}$, so no significant overlap with our $c \bar c$ Fock state is expected. Direct calculations for the molecular scenario are not available yet. Here, the neutral $D, \bar D^*$ mesons interact through their (transition) magnetic moments. One expects a much faster falloff at large $Q^2$ for these large-size objects. Future precise measurements in single tagged $e^+ e^-$ collisions can therefore offer valuable new insight into the structure of this enigmatic hadron and confirm or rule out an essential role of the $c \bar c$ component.
One may also explore the Primakoff--production in the field of a heavy nucleus in high-energy electron-nucleus collisions at the future electron-ion collider. Complementary information can be obtained also for timelike $B_c \to \chi_{c1}(3872)$ form factors, see Ref.~\cite{Colangelo:2022awx}.\\

\textit{Acknowledgements}--
This work was partially supported by the Polish National Science Center grant UMO-2018/31/B/ST2/03537 and by the Center for Innovation and Transfer of Natural Sciences and Engineering Knowledge in Rzesz{\'o}w. R.P.~is supported in part by the Swedish Research Council grant, contract number 2016-05996, as well as by the European Research Council (ERC) under the European Union's Horizon 2020 research and innovation programme (grant agreement No 668679).

\bibliography{bib.bib}
\end{document}